\documentstyle[prd,aps,preprint,epsfig,floats]{revtex}
\tightenlines

\input epsf.tex
\def\DESepsf(#1 width #2){\epsfxsize=#2 \epsfbox{#1}}

\begin{document}

\draft

\preprint{\vbox{
\hbox{OSU-HEP-02-14}
\hbox{UMD-PP-03-026}
}}

\title{{\Large\bf
 Lepton Flavor Violation and the Origin of}\\[0.07in]
{\Large\bf  the Seesaw Mechanism}}
\author{{\bf K.S. Babu}$^{1}$, {\bf B. Dutta}$^2$ and {\bf R.N. Mohapatra}$^3$ }

\address{
$^1$Department of Physics, Oklahoma State University,  Stillwater, OK
74078, USA\\
$^2$Department of Physics, University of Regina, SK S4S 0A2, Canada \\
$^3$Department of Physics, University of Maryland, College Park, MD 20742, USA}
\maketitle

\begin{abstract}

The right--handed neutrino mass matrix that is central to the understanding
of small neutrino masses via the seesaw mechanism can arise either
(i) from  renormalizable operators  or (ii) from nonrenormalizable or
super-renormalizable operators, depending on the symmetries and the
Higgs content of the theory beyond the Standard Model. In this paper, we
study lepton flavor violating (LFV) effects
in the first class of seesaw models wherein
the $\nu_R$ Majorana masses arise from renormalizable Yukawa couplings involving
a $B-L = 2$ Higgs field. We present detailed predictions for
$\tau \rightarrow \mu + \gamma$ and $\mu \rightarrow e + \gamma$ branching ratios
in these models taking the current neutrino oscillation data into account.
Focusing on minimal supergravity
models, we find that for a large range of MSSM parameters suggested by the relic
abundance of neutralino dark matter and that is consistent with Higgs boson mass
and other constraints, these radiative decays are in the range accessible to
planned experiments.  We compare these predictions with lepton flavor violation
in the second class of models arising entirely from the Dirac Yukawa couplings.
We study the dependence of the ratio $r \equiv B(\mu \rightarrow
e+\gamma)/B(\tau\rightarrow\mu +\gamma)$ on the MSSM parameters
and show that measurement of $r$ can provide crucial insight into the origin of the
seesaw mechanism.

\end{abstract}

\vskip1.0in

\newpage

\section{Introduction}

With the evidence for neutrino masses and mixings from solar and
atmospheric neutrino data getting more and more firm, the nature of new
physics that could explain the observations is under a great deal of
scrutiny \cite{caldwell}. There are several issues that need to be
understood, notably (a) the smallness of neutrino masses, and (b) the
origin of the large atmospheric neutrino as well
as the solar neutrino mixing angles, the latter being favored\cite{bahcall}
by the combined
solar neutrino results\cite{solar} including the recent SNO neutral current data
\cite{sno}.

Seesaw mechanism\cite{seesaw} provides one of the simplest ways to
understand the small neutrino masses.  It assumes the existence of
a set of three right--handed neutrinos ($\nu_R$) that have masses at
the scale $v_{B-L}$ corresponding to a new (local) $B-L$ symmetry of weak
interactions. Atmospheric neutrino observations suggest that the scale
$v_{B-L}$ is much lower than the Planck scale, leading to a new
threshold inferred solely from experiments. At low energies, these heavy
right--handed neutrinos induce operators of the form  $Y^2_{\nu}(LH_u)^2/v_{B-L}$, where
$Y_{\nu}$ is the Dirac Yukawa coupling matrix connecting the $\nu_R$ with the
left-handed neutrinos ($\nu_L$)  and the Standard Model (SM) Higgs doublet $H_u$.
$L$ denotes the lepton doublet which contains $\nu_L$.
After electroweak symmetry breaking, these operators induce masses for
the light $\nu_L$ of order $Y^2_{\nu}v^2_{wk}/v_{B-L}$.  Since $v_{B-L}$ is
much larger than the weak scale $v_{wk}$, the induced $\nu_L$
masses will be extremely small compared to
all other masses of the SM fermions.  Present neutrino oscillation data
suggests a scale $v_{B-L} \sim 10^{12}-10^{15}$ GeV.

The precise seesaw formula for the light neutrino mass matrix
with three generations  is given by
\begin{eqnarray}
{\cal M}_{\nu} = - M_D^T M^{-1}_{R}M_D~,
\end{eqnarray}
where $M_D$ is the Dirac neutrino mass matrix and $M_R$ is the $\nu_R$ Majorana
matrix. The neutrino mixing angles in such schemes
would arise as a joint effect from two
sources: (i) mixings among the right--handed neutrinos present in $M_{R}$
and (ii) mixings among different generations present in the Dirac mass
matrix $M_D$.  The (physical) neutrino oscillation angles will also receive
contributions from mixings among the charged leptons.

\subsection{Neutrino mixings and lepton flavor violation}

The neutrino flavor mixings induced by the seesaw mechanism
(as needed by oscillation data) can lead to lepton flavor violating
(LFV) effects.  However, within the SM extended minimally to accommodate the
seesaw mechanism, such effects are extremely small in any process other than
neutrino oscillation itself, owing to a power suppression factor
$(1/v_{B-L})^2$, as required by the decoupling theorem.  This situation
is drastically different if there is low
energy supersymmetry. The power suppression then becomes $(1/M_{SUSY})^2$
which is very much weaker and can lead to observable LFV effects at low
energies, as noted in a number of papers\cite{lfv1,lfv2,lfv3}.  The main
difference is that with low energy SUSY, lepton flavor violation can
be induced in the slepton sector, which can then be transferred to the
leptons suppressed only by a factor $(1/M_{SUSY})^2$. The way this
comes about is as follows.  Consider the minimal $N=1$ supergravity (mSUGRA)
models \cite{nath}. At the
fundamental scale $M_{Pl}$ where SUSY breaking is communicated to the SM sector,
all the soft SUSY breaking scalar masses are taken to be universal, and the
trilinear $A$--matrices are taken  to be proportional to the respective Yukawa
matrices.  Thus, at the Planck scale, there is no flavor violation anywhere
except in the Yukawa couplings and SUSY--preserving mass terms in the
superpotential.  Now, the right--handed neutrinos have masses
of order $v_{B-L}$ which is much lower than $M_{Pl}$, as inferred
from atmospheric neutrino data.  In the momentum regime
$v_{B-L} \leq \mu \leq M_{Pl}$ where the $\nu_R$ fields are active,
the soft masses of the sleptons will feel the effects
of LFV in the neutrino Yukawa sector through the renormalization group evolution.
At the scale $v_{B-L}$, the slepton mass matrix is no longer universal, and
this non-universality will remain down to the weak scale.  This LFV in the slepton
sector is subsequently
transferred to the leptons through one--loop diagrams involving the exchange
of gauginos.

It should be noted that
in the absence of neutrino masses, there is only one leptonic
Yukawa matrix, $Y_\ell$ for the charged leptons, which
can be diagonalized at $M_{Pl}$.  $Y_\ell$ will remain diagonal to the
weak scale, and would not induce any flavor violation in the slepton sector
in mSUGRA models.  The experimental evidence for neutrino oscillation is
thus a strong indicator that there might very well be lepton flavor violation,
assuming the validity of low energy SUSY.
Searches for LFV processes such as $\tau\rightarrow \mu +\gamma$ and/or
$\mu\rightarrow e +\gamma$ can therefore be an important source of
information on the $\nu_R$ mixings in $M_R$ and/or
family mixings in $M_D$.  (There are some exceptional
circumstances, which we point out later in this section.)

\subsection{Majorana LFV versus Dirac LFV}

It is possible to classify the seesaw models into two classes
depending on the way the $\nu_R$ gets its Majorana mass $M_R$. In
one case, $M_R$ arises from renormalizable Yukawa couplings
involving a $B-L = 2$ Higgs field $\Delta$ through
the superpotential terms
\begin{equation}
W \supset f \nu^c
\nu^c \Delta + Y_\nu \nu^c LH_u~.
\end{equation}
Here $f$ is the Majorana Yukawa coupling matrix, $Y_\nu$ is the Dirac
Yukawa matrix and
we use the standard notation appropriate
for supersymmetry with $\nu^c$ being the conjugate of $\nu_R$.  In the
second class of models, the matrix $M_R$ is either put in ``by hand'' as a
bare mass term in the Lagrangian (such terms are
super-renormalizable), or arises from non-renormalizable operators
involving a $B-L=1$ Higgs boson $\chi^c$ via the couplings
$(\nu^c\chi^c)^2/M$.   In models of the second class, the flavor violating
slepton mixings can arise (assuming mSUGRA) only from the Dirac Yukawa
couplings of the
neutrinos, $Y_\nu$.
On the other hand, in the first class, which is the main focus of
this paper, LFV can arise from both  the Dirac Yukawa coupling
$Y_\nu$ as well as the Majorana Yukawa coupling $f$.

For the purpose of the present paper,
we will call the two lepton flavor violation alternatives as
LFV in the Majorana case and
LFV in the Dirac case.

The simplest example of models with renormalizable Majorana Yukawa
couplings is provided by the minimal (SUSY) extension of the SM with a gauged
$B-L$ symmetry.  Bare masses for the $\nu_R$ are then forbidden by $B-L$
gauge invariance.  If the Higgs sector contains a $B-L = 2$ field
$\Delta$, the Yukawa coupling $f \nu^c \nu^c \Delta$ as in Eq. (2)
will be allowed
which will induce $M_R$ of order $v_{B-L} \equiv \left\langle \Delta \right
\rangle$.  (If instead of $\Delta$, a $B-L=1$ Higgs field $\chi^c$ is
used, $M_R$ can arise from a non-renormalizable coupling $(\nu^c \chi^c)^2/M$.
This belongs to the second class.)
There are interesting extensions of this class of models with renormalizable
Majorana couplings, such as SO(10) with a {\bf 126} Higgs field\cite{babu} or
left-right symmetric model with a triplet Higgs field.
In these models the Majorana Yukawa couplings may play
a dominant role in generating the required neutrino mixings.  The role
of the Dirac Yukawa coupling $Y_\nu$ for LFV may be subleading and
may even be negligible.  To be concrete, we shall assume in our
analysis that the matrix $Y_\nu$
is diagonal in a basis where the charged lepton Yukawa matrix $Y_\ell$
is also diagonal. For numerical purposes we shall further assume that
$Y_\nu$ and $Y_\ell$
are proportional, a situation realized naturally if the gauge symmetry is left-right
symmetric.
The neutrino mixings will all arise from the Majorana coupling matrix $f$, which
will be determined (upto an overall scale factor) from neutrino oscillation data.
The proportionality assumption $Y_\nu \propto Y_\ell$
is not crucial to our main conclusions on LFV effects
related to $M_R$, but it greatly simplifies the presentation of our results.
This assumption may be viewed as analogous to the assumption often made
in the case of  Dirac LFV  models that the super-renormalizable Majorana
mass matrix $M_R$ is proportional to an identity matrix.

\subsection{Summary of results}

In an earlier paper \cite{bdmlfv} we have presented results of our preliminary
investigations on LFV involving the Majorana Yukawa matrix $f$, assuming
proportionality of $Y_\nu$ and $Y_\ell$.  (We called this
relation up--down unification.)  There we studied a general
class of SUSY breaking models and did not adhere  to mSUGRA.  The analysis
of Ref. \cite{bdmlfv} was carried out in the context of left--right symmetric
gauge theories with an enhanced gauge structure.  In contrast to Ref. \cite{bdmlfv},
in this paper we analyze LFV effects induced by the $f$ matrix
within mSUGRA models.  This means that the
only source of lepton flavor violation is in the neutrino Yukawa sector.  Significant
LFV effects are found even in this minimal scenario.  As explained in Sec. II.C,
a novel way of fitting the low energy neutrino oscillation data has the
effect of enhancing the LFV in this minimal scheme.  We
also use a simplified gauge structure involving only an extra $B-L$
symmetry.  Even this symmetry is not essential, the crucial ingredient is
the Yukawa coupling matrix $f$ of Eq. (2).  Thus the models studied here
are more minimal compared to the left-right models and are
likely to be the low energy limits of a wider class of unified models.

Our main result is that LFV effects associated with the Majorana Yukawa
coupling $f$ are in the range accessible to forthcoming experiments
for a large range of MSSM parameters
consistent with constraints from dark matter, the Higgs
boson mass and direct experimental search for supersymmetry.
We also point out that the detailed dependence
of the branching ratios in processes such as
$\tau \rightarrow \mu+ \gamma$ and $\mu \rightarrow e + \gamma$
on the MSSM parameters is different in this class of models compared
to the case of flavor violation associated with the Dirac Yukawa coupling
$Y_\nu$.  With some information of the MSSM parameters, measurements
of these radiative decays will enable one to pin down
the nature of
interactions that give rise to the right-handed neutrino masses. Obviously, this
will have a profound impact on our  understanding of the neutrino
sector and can shed light on the origin of the seesaw mechanism.

It may be worth pointing out that there are other
experimental predictions that may help to distinguish between these two
possibilities. For instance, two of us  have recently
pointed out that the Dirac LFV models  can lead to an observable signal in the
baryon number  non-conserving process neutron-anti-neutron
oscillation\cite{nnbar} whereas in the Majorana LFV case, the N-$\bar{N}$
oscillation is unobservable.

\subsection{One caveat}

There is one caveat in the foregoing discussions that should
be mentioned.  It applies equally
well to the Dirac and Majorana LFV alternatives.  In a special class of
seesaw schemes it  is possible that
no lepton flavor violation  shows up in the next round of
experiments without contradicting the current neutrino oscillation data
even in the presence of low energy SUSY.
While we feel that such  situations would be
non--generic, it should be born in mind, nevertheless.  In the Dirac LFV
alternative it occurs as follows.
It is conceivable that all the flavor mixings arise from
the mass matrix $M_R$ and that $Y_\nu$ is diagonal. Flavor violation
in the bare mass parameter (or non-renormalizable term) $M_R$ is
not transmitted to the slepton sector via the RGE, implying highly
suppressed LFV effects. One can easily fit
all the neutrino oscillation data via the mixings present in $M_R$.
It is also possible that the scale $M_R$ is much below $10^{12}$
GeV, in which case the elements of $Y_\nu$ will  be much smaller
than unity so as  to fit the neutrino data, again suppressing LFV effects.
Similarly, in the Majorana LFV alternative, it  might so happen that
the elements of $f$ are very small, compensated by a higher value of $v_{B-L}$ so
that $M_R$ is unchanged, or that the elements of $Y_\nu$ are small.
Such a scenario will suppress lepton flavor violation.
A null result in the rare lepton decay
searches would point to a situation in which either of these two
situations could be operative and we will have no way to tell which
operators are inducing  the $\nu_R$ masses.  On the other hand, if
there is a positive signal in the rare
decays, our results will help in probing the origin of $M_R$.
It appears
to us natural to have at least the third family Yukawa coupling in $Y_\nu$
of order one (analogous to the top quark Yukawa coupling being of order one)
 and similarly at least one element of $f$
($f_{33}$) to be  of order one.  Observable LFV effects follow in this
case, which is what we analyze.

\subsection{Outline of the paper}

We have organized this paper as follows. In Section II, we present the
basic model and the relevant renormalization group equations (RGE)
 that will be used in
getting the slepton mixings and thus
the lepton flavor violating rates for the
case of Majorana LFV.  Here we also present analytic approximations to
the LFV effects by integrating the RGE.  These expressions, while
only approximate, are useful in gauging the
significance of LFV in the Majorana case as well as in comparing it with
the case of Dirac LFV.  In Section II.A we present the relevant
expressions for obtaining the $f$ matrix by inverting the seesaw formula.
In II.B this information is used to compare analytically the predictions for
$\tau \rightarrow \mu + \gamma$ and $\mu \rightarrow e + \gamma$ branching
ratios in the two cases.  In Section II.C we present a convenient way
to parametrize the Majorana Yukawa matrix $f$ that can be compared directly
against low energy neutrino oscillation data on the one hand and LFV effects
on the other.  Section III is devoted to our exact numerical results for
LFV effects in the case of Majorana as well as Dirac Yukawa couplings.  In
III.A we present our numerical fits to the neutrino parameters, III.B has
our results for the branching ratios $\tau \rightarrow \mu+ \gamma$ and
$\mu \rightarrow e+\gamma$ as well as the ratio of the two branching ratios.
Section IV has our conclusions.

\section{Slepton mixing and lepton flavor violation from Majorana Yukawa
Couplings}

In order to illustrate in detail the source of LFV for the case of
Majorana Yukawa couplings generating the $\nu_R$ mass matrix (the Majorana LFV
alternative), we work with
a simple extension of the minimal supersymmetric standard model that
accommodates the seesaw mechanism.  It is
based on
the weak gauge group $SU(2)_L\times U(1)_{I_{3R}}\times U(1)_{B-L}$ with
fermions assigned as follows: $Q(2, 0, +\frac{1}{3});~ L(2, 0, -1);~ u^c
(1,-\frac{1}{2}, -\frac{1}{3});~ d^c (1, +\frac{1}{2}, -\frac{1}{3});~ e^c
(1, +\frac{1}{2}, +1);~ \nu^c (1, -\frac{1}{2}, +1)$.  This is
the simplest extension of the weak interaction sector
 that predicts the existence of the $\nu_R$
from gauge anomaly cancellations conditions and thus to small neutrino
masses via the seesaw mechanism.  The Higgs superfields
are $H_u (2, +\frac{1}{2}, 0);~ H_d (2, -\frac{1}{2}, 0);~ \Delta (1, +1,
-2),~ \bar{\Delta} (1, -1, +2)$.
The superpotential for this theory is:
\begin{eqnarray}
W&=&Y_u u^c QH_u + Y_d d^c Q H_d + Y_{\ell} e^c L H_d + Y_{\nu} \nu^c LH_u\\
\nonumber
&+&f\nu^c\nu^c\Delta + \mu H_uH_d + S(\Delta\bar{\Delta}-M^2)~.
\end{eqnarray}
Here all the Yukawa couplings are $3\times 3$ matrices; $S$ is a singlet
field introduced to help facilitate the breaking of
$U(1)_{B-L}\times U(1)_{I_{3R}}$ down to $U(1)_Y$ in the SUSY limit.
One identifies the SM hypercharge as $Y/2 = I_{3R} + (B-L)/2$.
This
superpotential gives
equal vacuum expectation values (VEVs) to $\left\langle\Delta\right\rangle
=\left\langle\bar{\Delta}\right\rangle =
v_{B-L}$.  The Yukawa coupling matrix $f$ will induce large mass for the
$\nu^c$ field, leading to the seesaw mechanism.
This model should be contrasted with the commonly discussed case in the
literature where the heavy $\nu_R$ masses are either put in by hand or
arise from a non-renormalizable term in the superpotential.

In order to study flavor violation, we will make the simplifying
assumption that both $Y_{\nu}$ and $Y_\ell$ are diagonal at the Planck
scale; therefore all neutrino mixings arise from the general flavor
structure of the Majorana Yukawa couplings $f_{ij}$.
Then from the
approximate bi-maximal pattern for neutrino mixings deduced from
oscillation data, we can obtain $f_{ij}$ for different patterns of
neutrino masses. Since the model is supersymmetric, the flavor violation present in
$f_{ij}$ will induce processes such as
$\tau\rightarrow \mu + \gamma$ and $\mu\rightarrow e +\gamma$ decays. We
investigate these predictions using simple analytic approximations in
this section.

We need to write down the RGE for the Yukawa couplings and slepton masses
for the class of models under consideration.  Denote the left--handed slepton
mass squared matrix as $m_L^2$ and the trilinear $A$ terms
in the slepton sector as $A_\ell,~ A_\nu,
~A_f$ (in an obvious notation). The relevant RGE for $Y_\nu$ and $m_L^2$ are:
\begin{eqnarray}
\frac{dY_{\nu}}{dt}~=~\frac{Y_{\nu}}{16\pi^2}
[Tr(3Y_uY^{\dagger}_u+Y_{\nu}Y^{\dagger}_{\nu}) +3Y^{\dagger}_{\nu}Y_{\nu}
 +Y^{\dagger}_{\ell}Y_{\ell}+4f^{\dagger}f -3g^2_2-
g^2_{R}-{3\over 2}g^2_{B-L}] \\
\frac{dm^2_L}{dt}~=~\frac{1}{16\pi^2}[ (m^2_L + 2m^2_{H_d})
Y^{\dagger}_\ell Y_\ell
+(m^2_L+2m_{H_u}^2)Y^{\dagger}_{\nu}Y_{\nu} +2Y^{\dagger}_\ell m^2_{e^c}Y_\ell
+Y^{\dagger}_\ell Y_\ell m^2_L\\ \nonumber+2Y^{\dagger}_\nu
m^2_{\nu^c}Y_\nu
+Y^{\dagger}_\nu Y_\nu m^2_L
 +2A^{\dagger}_\ell A_\ell +2A^{\dagger}_\nu A_\nu
 -6g^2_2M_2^2-3g^2_{B-L}M_{B-L}^2]
\end{eqnarray}
The RGE for the relevant $A$ parameters are:
\begin{eqnarray}
\frac{dA_\ell}{dt}&=& \frac{1}{16\pi^2} [A_\ell [ Tr (3Y^{\dagger}_dY_d +
Y^{\dagger}_\ell Y_\ell) + 5 Y^{\dagger}_\ell Y_\ell+Y^{\dagger}_\nu Y_\nu
\\ \nonumber
&-&3g^2_2-g^2_{R}-{3\over 2} g^2_{B-L}] +Y_\ell[Tr(6A_dY^{\dagger}_d+2 A_\ell
Y^{\dagger}_\ell)\\ \nonumber &+& 4 Y^{\dagger}_\ell A_\ell+
2Y^{\dagger}_\nu A_\nu+
6g^2_2M_2 +2g^2_{R}M_{R}+3g^2_{B-L}M_{B-L}]]
\end{eqnarray}
\begin{eqnarray}
\frac{dA_\nu}{dt}&=& \frac{1}{16\pi^2} [A_\nu [ Tr (3Y^{\dagger}_uY_u +
Y^{\dagger}_\nu Y_\nu) + 5 Y^{\dagger}_\nu Y_\nu+Y^{\dagger}_\ell Y_\ell
\\ \nonumber
&+& 4f^{\dagger}f-3g^2_2-g^2_{R}-{3 \over 2}g^2_{B-L}]
+Y_\nu[Tr(6A_uY^{\dagger}_u+2 A_\nu Y^{\dagger}_\nu) \\ \nonumber&+&
4 Y^{\dagger}_\nu A_\nu+8f^{\dagger}A_f+2 Y^{\dagger}_\ell
A_\ell+6g^2_2M_2 +2g^2_{R}M_{R}+3g^2_{B-L}M_{B-L}]]
\end{eqnarray}
We are interested in LFV arising primarily from the matrix $f$. Adopting
mSUGRA,  we  have at the Planck scale,
$A_\ell = A_0 Y_\ell,~ A_\nu = A_0 Y_\nu,~A_f = A_0 f$.  Furthermore,
at $M_{Pl}$, we have universality of all scalar masses (denoted by
$m_0$) and a common
gaugino mass (denoted as $m_{1/2}$).

Under the set of assumption specified above,
it is clear from Eqs. (4)-(7) that the off--diagonal elements
of $f$ will not induce slepton flavor violation to the lowest order
in a perturbative expansion, since $f$ does not enter directly in
the RGE for $m_L^2$ or $A_\ell$.
However, the one--loop solution for
$m^2_L$ involves $Y^{\dagger}_{\nu}Y_{\nu}$, which
receives a contribution from $f^{\dagger}f$ (see Eq. (4)). Therefore
in second order in the parameter $\frac{1}{16\pi^2}\ell n
\left(\frac{M_{P\ell}}{M_{B-L}}\right)$, we do have flavor violating
effects in $m_L^2$ and $A_\ell$. It is important to note that
this is not a two--loop RGE effect, rather that it is a one--loop RGE
improved effect, as indicated by the presence of the
[$\ell n\left(\frac{M_{P\ell}}{M_{B-L}}\right)]^2$ term.

We can estimate the
strength of the slepton flavor violation by integrating the RGE analytically.
We find from Eq. (4)-(7) that
\begin{eqnarray}
\Delta m^2_{ij}(i\neq j)~\simeq ~
\frac{-3(m^2_0+A_0^2)}{32\pi^4}[Y^{\dagger}_{\nu}Y_{\nu}f^{\dagger}f
+f^\dagger f Y_\nu^\dagger Y_\nu]_{ij}\left(\ell
n\frac{M_{P\ell}}{M_{B-L}}\right)^2
\end{eqnarray}
\begin{eqnarray}
A_{\ell ij}(i\neq j)~\simeq ~
\frac{-3}{64\pi^4}[A_\ell(Y^{\dagger}_{\nu}Y_{\nu}f^{\dagger}f
+ f^\dagger f Y_\nu^\dagger Y_\nu)]_{ij}\left(\ell
n\frac{M_{P\ell}}{M_{B-L}}\right)^2~.
\end{eqnarray}
The branching ratio for $\ell_j\rightarrow \ell_i + \gamma$ can be written
approximately as
\begin{eqnarray}
{B(\ell_j\rightarrow \ell_i +\gamma) \over B(\ell_j \rightarrow \ell_i +\nu_j
+ \bar{\nu}_i)}
\simeq \frac{3\alpha_{em}(c_1g^2_1 +
g^2_2)^2}{32\pi m^4_{sl}G^2_F}
\left({\Delta m^2_{ij} \over m_{sl}^2 }\right)^2\tan^2\beta~.
\end{eqnarray}
Here $m_{sl}^2$ is the average slepton mass at the weak scale, which may
be quite different from $m_0^2$.  Eq. (10) is obtained as an approximation
to the chargio and neutralino exchange diagram proportional to LFV in Eq. (8)
with the enhancement factor
of $\tan\beta$ as indicated. $c_1$ in Eq. (10) is an ${\cal O}(1)$ coefficient.
LFV arising from Eq. (9) can be comparable,
but it will not change our approximate results by much.
We have found that Eq. (10) is a reasonable
approximation to the exact numerical results to within a factor of few,
provided that $\tan\beta$ is not
too large and that there is no large hierarchy between the SUSY breaking
mass parameters ($\mu,~m_{sl},~m_{1/2}$).

Within mSUGRA, $m_{1/2}$ must be larger than about 250 GeV, to be consistent
with the Higgs boson mass constraint (viz., $m_h \ge 114$ GeV) and the $b\rightarrow s\gamma$
constraint.
Since we also want to identify the
lightest neutralino (LSP) as the dark matter in the universe, the relic abundance
puts another constraint that the second lightest SUSY particle (NLSP), the
right--handed stau in our
case, should be nearly degenerate with the LSP to within about 20-30 GeV.
This condition can be
satisfied if $m_{1/2} \simeq 4.4 m_0$ is obeyed, which we shall adopt.
Take for example a relatively light SUSY spectrum with
$m_{1/2} = 250$ GeV and $\tan\beta = 10$.  The charged
Wino mass is then approximately 200 GeV, and the Bino mass is about 100 GeV.
Choosing $m_0 = 55$ GeV and $A_0 = 0$ leads to a mass of about 190 GeV for
the left--handed
sleptons and a mass of about 130 GeV for the right--handed sleptons, with
the right--handed
stau being somewhat lighter around 115 GeV.  Co-annihilation of LSP will
be efficient with such a spectrum.

Next we need to choose the neutrino parameters.  With $Y_\nu \propto Y_\ell$,
a good fit to the current
neutrino oscillation data can be obtained with the choice $(Y_\nu)_{33} = 0.66$
and $(f^\dagger f)_{23} \simeq 0.28$ (see detailed numerical fits later).
Furthermore, $\ell n(M_{Pl}/M_{B-L}) \simeq 9$ is quite plausible.  With
this choice, we estimate $B(\tau \rightarrow \mu + \gamma) \simeq
2 \times 10^{-9}$ for $\tan\beta = 10$ from Eq. (10).  As $A_0$ is increased
from 0 to 300 GeV, this branching ratio increases to about
$2 \times 10^{-6}$.  Thus we see that for interesting ranges of parameters,
the process $\tau \rightarrow \mu+\gamma$ can be within experimental reach.
More careful examination of the decay is thus warranted.  Similarly, the
decay $\mu \rightarrow e + \gamma$ can have a branching ratio
 of order $10^{-14}-10^{-9}$
for reasonable ranges of model parameters.  Some of this parameter space is
already excluded by data, there is a good chance of being able to measure
this decay in the planned experiments.

It is instructive to compare the branching ratios obtained here in the case of
Majorana LFV to that expected in the case of Dirac LFV.  In the case of
Dirac LFV, assuming that the $\nu_R$ Majorana mass matrix is proportional
to the identity matrix, we can write down the analog of Eq. (8) as
\begin{equation}
\Delta m^2_{ij}(i\neq j) \simeq -{1 \over 8 \pi^2}(3m_0^2+A_0^2)(Y_\nu^\dagger
Y_\nu)_{ij} \left(\ell n {M_{Pl} \over M_{B-L}}\right)~.
\end{equation}
Comparing Eq. (11) with Eq. (8), we find that numerically the two effects
can be comparable.  Which effect is more dominant will
depend on the overall strengths of the coupling matrices in the two cases.
This will be discussed further in Sec. III.

\subsection{Flavor structure of the Majorana-Yukawa couplings}

We can invert the seesaw formula (Eq. (1)) and
obtain the matrix elements of the Majorana Yukawa coupling
matrix $f$:
\begin{eqnarray}
f~=~\frac{1}{v_{B-L}}M_D {\cal M}^{-1}_{\nu}M_D^T~.
\end{eqnarray}
We also have the relation for the inverse of the light neutrino mass matrix
${\cal M}_\nu^{-1}$ in terms of the light neutrino mass eigenvalues
$(m_1,~m_2,~m_3)$ and the leptonic mixing matrix ${\bf \Large U}$ as
\begin{eqnarray}
{\cal M}_\nu^{-1} =
{\bf \Large U}\left(\begin{array}{ccc} m^{-1}_1 & 0 & 0 \\ 0 & m^{-1}_2 &
0 \\ 0 & 0 & m^{-1}_3\end{array}\right){\bf \Large U^T}~.
\end{eqnarray}
We ignore
CP violation and use the near bi-maximal
approximation for ${\bf \Large U}$  given by
\begin{eqnarray}
{\bf \Large U}~\simeq~\left(\begin{array}{ccc} c & s & \epsilon \\
-\frac{s+c\epsilon}{\sqrt{2}} & \frac{c-s\epsilon}{\sqrt{2}} &
\frac{1}{\sqrt{2}} \\
\frac{s-c\epsilon}{\sqrt{2}} & \frac{-c-s\epsilon}{\sqrt{2}} &
\frac{1}{\sqrt{2}} \end{array}\right)~.
\end{eqnarray}
Here $U_{e3} \simeq \epsilon \leq 0.16$ from the CHOOZ and Palo-Verde
reactor experiments\cite{chooz}. $s$ is the solar neutrino mixing angle
for which the allowed range after the SNO data is \cite{bahcall}
$0.71\leq {\rm sin}^22\theta_{\odot} \leq 0.94 $ at three $\sigma$
confidence level. Multiplying all these together, we have
\begin{eqnarray}
{M}_{R, ij}~=~ m_{D,i}\mu^{-1}_{ij}m_{D,j}
\end{eqnarray}
where we have assumed $M_D$ to be diagonal with its entries denoted by
$m_{D,i}$.  The definition $\mu_{ij}^{-1} \equiv ({\cal M}_\nu^{-1})_{ij}$ has been
used.   We have then
\begin{eqnarray}
\mu^{-1}_{11} &=& \frac{c^2}{m_1}+\frac{s^2}{m_2}+\frac{\epsilon^2}{m_3} \\
\nonumber
\mu^{-1}_{12} &=& -\frac{c(s+ c\epsilon)}{\sqrt{2}m_1} +\frac{s(c-
s\epsilon)}{\sqrt{2}m_2} +\frac{\epsilon}{\sqrt{2}m_3}
\\ \nonumber
\mu^{-1}_{13} &=& \frac{c(s- c\epsilon)}{\sqrt{2}m_1} - \frac{s(c+
s\epsilon)}{\sqrt{2}m_2} +\frac{\epsilon}{\sqrt{2}m_3} \\ \nonumber
\mu^{-1}_{22} &=& \frac{(s+c\epsilon)^2}{{2}m_1} +\frac{(c-
s\epsilon)^2}{{2}m_2} + \frac{1}{2m_3} \\ \nonumber
\mu^{-1}_{23} &=& -\frac{(s^2- c^2\epsilon^2)}{{2}m_1}-\frac{(c^2-
s^2\epsilon^2)}{{2}m_2} +\frac{1}{2m_3} \\ \nonumber
\mu^{-1}_{33} &=& \frac{(s- c\epsilon)^2}{{2}m_1}+\frac{(c+
s\epsilon)^2}{{2}m_2} +\frac{1}{2m_3}.
\end{eqnarray}
Using these expressions, for different scenarios of neutrino masses
we can determine the
values of $f_{ij}$ which enter into the flavor changing slepton masses.
We will focus primarily on the hierarchical neutrino mass schemes where
$m_1 \ll m_2 \ll m_3$.

\subsection{Comparison between Majorana LFV and Dirac LFV alternatives}

Let us first discuss the case where the right--handed neutrino masses arise
from a renormalizable Majorana Yukawa coupling $f$. As in the previous
section, let us assume that the Dirac neutrino mass matrix $M_D$ is diagonal
in the same basis where $M_\ell$ is diagonal.
One interesting pattern is provided by the
normal hierarchy $m_1 \ll m_2 \ll m_3$.  In this case, it is plausible,
but not mandatory,
that in Eq. (16), the terms with $m_1$ in the denominator dominate $\mu_{ij}^{-1}$
with all other terms being negligible.  In this approximation, we obtain
for the process $\tau \rightarrow \mu + \gamma$ process an amplitude
proportional to
\begin{equation}
(\Delta m^2)_{23} \propto {m_{D,2}m_{D,3}^5 s^4 \over 4 m_1^2 w_{wk}^2 v^2_{B-L}}~.
\end{equation}
Similarly, the $\mu \rightarrow e + \gamma$ process has an amplitude
given by
\begin{equation}
(\Delta m^2)_{12} \propto {m_{D,1}m_{D,2}^5 s^3 c \over 2 \sqrt{2}
 m_1^2 w_{wk}^2 v^2_{B-L}}~.
\end{equation}
We have set $\epsilon = 0$ for simplicity here.  From Eq. (17)-(18) it
might appear that the branching ratios can increase without limit
as $m_1$ is decreased.  However, this is not true.
The coupling $f_{33} \simeq m_{D,3}^2 s^2/(2m_1 v_{B-L})$ should remain
perturbative ($f_{33} \leq 1$).
This means that for a given $m_1$, there is a lower
limit on $v_{B-L}$.  For example, if $m_1 = 10^{-3}$ eV, and for $m_{D,3}
= 115$ GeV, $s = 0.52$, it must be that $v_{B-L} \ge 1.8 \times
10^{15}$ GeV so that $f_{33} < 1$.  It turns out that with this
constraint, the branching ratios (BR) for the decays
$\tau \rightarrow \mu+\gamma$ and $\mu \rightarrow e+\gamma$
 are suppressed within mSUGRA, mainly because
the logarithm upon which the BR has a fourth power dependence, is
small.  We conclude that in this scenarios with $1/m_1$ dominance in Eq. (16),
in the case of Majorana LFV there is no significant LFV in lepton
sector within mSUGRA.

We have also examined LFV effects if the mSUGRA assumption is relaxed
somewhat for the case of $1/m_1$ dominance.  If for example, $A_f= A_0'f,
~A_\ell = A_0 Y_\ell,~A_\nu = A_0 Y_\nu$ at $M_{Pl}$ with $A_0'$ different
from $A_0$, we  found that the BR for $\tau \rightarrow \mu+\gamma$ can
be in the range $10^{-7}-10^{-9}$ for relative large $A_0' \sim$ TeV.
The BR for $\mu \rightarrow e+\gamma$ is extremely small ($\sim 10^{-16})$
with this set of boundary conditions.  We shall not present details of this
fit, and will focus solely on the mSUGRA case.

For comparison,
let us also consider the case of Dirac LFV.  For $M_R \propto {\bf 1}$
there is an unambiguous connection between the LFV effects and
the neutrino mass matrix. To see this more explicitly, note that in this
case, the seesaw formula reduces to
\begin{eqnarray}
-M {\cal M}_{\nu} = M_D^TM_D~.
\label{dirac}\end{eqnarray}
Thus the product  $M_D^TM_D$ can be
completely expressed in
terms of the neutrino masses and mixings.
Since for the CP conserving case, the flavor violating slepton mixings
effects always involve $Y^T_\nu Y_\nu$, the LFV amplitudes can be
expressed directly in terms of the neutrino mass matrix.
Assuming again that
$m_1 \ll m_2 \ll m_3$, we can write down the amplitude for the decay
$\tau \rightarrow \mu + \gamma$ to be proportional to
$(\Delta m^2)_{23} \propto {m_3 \over 2}$, while that for the decay
$\mu \rightarrow e+\gamma$ is proportional to $(\Delta m^2)_{12}
\propto csm_2/\sqrt{2}$.  These amplitudes increase with the light
neutrino masses, unlike the case of Majorana LFV, which scale
inversely with $m_1$.  Both decays have branching ratios accessible
in the planned experiments in the Dirac LFV alternative, as will
be further explained in Sec. III.

\subsection{Enhanced LFV from Majorana Yukawa couplings}

The dominance of $1/m_1$ in $\mu^{-1}_{ij}$ and
therefore in $f_{ij}$ is not the only possibility.  We have found
a large range of parameters where the LFV effects are significant
while deviating from the $1/m_1$ dominance.
To see this explicitly it is more convenient to define the light neutrino mass
matrix ${\cal M}_\nu$ in the gauge basis as a perturbative expansion
in a small parameter $\epsilon \sim 0.1$
\begin{eqnarray}
{\cal M}_\nu = m_0 \left(\matrix{e \epsilon^n & h \epsilon^m & d\epsilon \cr
h \epsilon^m & 1+a \epsilon & 1 \cr d \epsilon & 1 & 1+b\epsilon}\right)~.
\end{eqnarray}
Here $(a,~b,~d,~e,~h)$ are order one coefficients, $\Delta m^2_{atm} \simeq 4
m_0^2$ and the exponents $(n,~m)$ in the (1,1) and (1,2) entries have to
be at least 1, but can be larger.
This matrix provides a good
fit to all neutrino data.  If $m=1$, and $n=1,~2$ it will reproduce the results of
the previous section with $1/m_1$ dominance.  But for other values of $(n,~m)$,
which are also consistent with all neutrino oscillation data, we have found much
larger LFV effects.  Consider the case
$n=2, m=4$ for example.  In this  case the (1,1) and
the (1,2) entries of ${\cal M}_\nu$ are not completely specified by
neutrino data alone.  Inverting Eq. (20) we obtain for the matrix $f$,
\begin{eqnarray}
f={m_{D,3}^2 \over d^2 m_0 v_{B-L}} \left(\matrix{(a+b)c^2 \epsilon^5 & cd \epsilon^3 & -cd
\epsilon^2 \cr cd \epsilon^3 & -d^2 \epsilon^2 & dh \epsilon^2 \cr
-cd \epsilon^2 & dh \epsilon^2 & (e-h^2)\epsilon^2}\right)~.
\end{eqnarray}
Here we have used the parametrization $M_D = {\rm Diag}(c \epsilon^3,~
\epsilon,~1)m_{D,3}$.
All parameters in $f$ are fixed, except for $(e,~h)$.  If $(e,~h) \sim {\cal O}(1)$,
we have a situation where the off--diagonal entries as well as the
diagonal entries of $f$ are of the same order leading to large LFV effects.
Contrast this with the case where
$m=1$ is adopted in Eq. (20), in which case we would have the (3,3) entry of $f$ to
be order 1, the (2,3) to be order $\epsilon$, the (1,3) of order $\epsilon^2$,
and the (1,2) entry of order $\epsilon^3$.
The choice (which is equivalent to $1/m_1$ dominance in $f$)
will lead to a suppression of the $\tau \rightarrow \mu+\gamma$ branching ratio
by a factor $\epsilon^2 \sim 10^{-2}$ which would put it in an
unobservable range.  The decay $\mu \rightarrow e+\gamma$ will be suppressed
even further, by a factor $\epsilon^4 \sim 10^{-5}$, compared to the case
of $m=2$.

It is clear then that the most interesting case for lepton flavor violation
is when $n=4,~m=2$ in Eq. (20).  This is the choice we shall make for our
numerical fits in Sec. III.

\section{Numerical Results}

\subsection{Neutrino mass fit}

We start with a basis where the charged lepton mass matrix is diagonal. For the
Dirac neutrino mass matrix we assume that
\begin{equation}
M_D =\gamma\tan\beta M_\ell~,
\end{equation}
where $M_\ell={\rm Diag}(m_e,\,m_{\mu},\,m_{\tau})$.
This assumption is well justified if the model is embedded into a SUSY
left--right framework as we have shown earlier papers\cite{bdm}.
It is also possible that this assumption holds approximately in a wider
class of models where
the off--diagonal entries in $Y_\ell$ and $Y_\nu$ are small, although
not strictly
proportional.  The proportionality constant $\gamma$ is a parameter of
the model.  If we assume that the $\nu_\tau$ Dirac Yukawa coupling is the
same as the top--quark Yukawa couplings, as happens in some versions of
minimal $SO(10)$ models,
then $\gamma = 1$ for large values of $\tan\beta \approx 60$ while
$\gamma  \simeq 5-6$ for $\tan\beta \approx 10$.

The light neutrino
Majorana mass matrix is obtained as
\begin{equation}
{\cal M}_{\nu}={{\gamma^2\tan^2\beta}\over v_{B-L}}M_\ell f^{-1}M_\ell~.
\end{equation}
 In our analysis, we choose values of $\tan\beta \geq 10$,
since smaller $\tan\beta$ values are less preferred by the
recent Higgs mass bounds from LEP \cite{LEP} and the muon $g-2$
anomaly\cite{bnl}, if interpreted as a SUSY effect, would prefer a moderately
large $\tan\beta$.

Following the analytic expression given in Sec. II.C, we obtain a fit
for the light neutrino spectrum as follows.
We choose tan$\beta$ = 10 and  find a good fit to the solar and atmospheric
neutrino data with hierarchical neutrino masses if the matrix  $f$
defined at at $v_{B-L} = 2\times
10^{12}$ GeV has entries given by
\begin{eqnarray} {f}&=&\left(\matrix{
  -1.1\times 10^{-4} & -0.015 & 0.29\cr
  -0.015 & 0.50 & -0.57\cr
 0.29& -0.57& 0.104  }\right).
\end{eqnarray}
The resulting neutrino masses at the weak scale (using the RGE
evolution \cite{babu2}) are:
\begin{equation}
(m_1,~m_2,~m_3) = (-2.7\times
10^{-3},\, 6.4\times 10^{-3},\,  8.6\times 10^{-2})~ {\rm eV}~.
\end{equation}
The leptonic mixing
matrix is given by:
\begin{eqnarray} {\Large\bf U}&=&\left(\matrix{
 0.85 & -0.52 & -0.053 \cr
  0.33& 0.62 & -0.72\cr
  -0.40& -0.59& -0.70}\right).
\end{eqnarray}
Note that the mixing angles $|U_{e2}| = 0.52$ and $|U_{\mu 3}| = 0.72$
are in very good agreement with the solar and atmospheric oscillation data.
 Note also the prediction for the parameter $|U_{e3}|\simeq 0.05$,
which is in the range that can be tested in planned long baseline
experiments.

\begin{figure}\vspace{-2cm}
\centerline{ \DESepsf(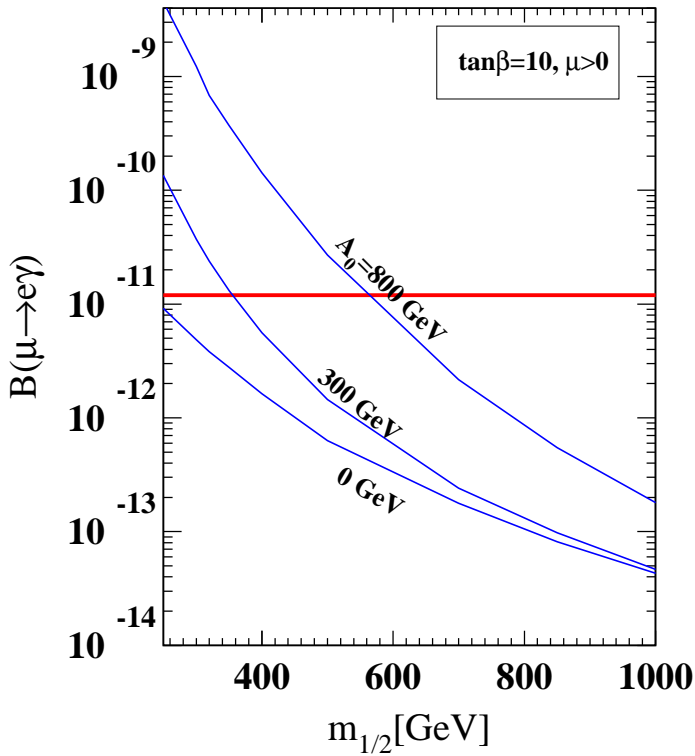 width 8 cm) }
\vspace{0.5cm}\caption {\label{fig1}$B(\mu\rightarrow
    e+\gamma)$ vs $m_{1/2}$ for three values of $A_0$
     corresponding to the case of Majorana LFV.
    The Yukawa coupling matrix $f$ is given in Eq. (24).  The horizontal
    solid line shows the current experimental limit.}
\end{figure}

For comparison purposes we will also present LFV effects associated
with Dirac Yukawa couplings, assuming no flavor violation in the
Majorana sector.  It will be
desirable to have a neutrino mass fit for this case as well,
where $M_R$ is proportional to an identity matrix and all flavor
mixings arise from $Y_\nu$.  We have found such fits for the same
set of light neutrino spectrum as in Eq. (25) -(26).  Actually, there
are discrete possibilities for $Y_\nu$
since the equations are not linear in $Y_\nu$.  We have found that
the predictions for LFV decays are essentially unchanged within these
discrete possibilities. So we focus on one of them.
At $v_{B-L} =9\times 10^{13}$ GeV, we find that $Y_\nu$ is given by
\begin{eqnarray} {Y_{\nu}}&=&\left(\matrix{0.04+ 0.074 i & -0.073+
      0.029i& 0.025 - 0.034 i\cr
  -0.073+  0.029 i & -0.22 + 0.011 i& -0.35- 0.013 i\cr
 0.025-0.034 i& -0.35 - 0.013 i& -0.24+0.016 i}\right).
\label{ynu}\end{eqnarray}
Although $Y_\nu$ in Eq. (27) have complex elements, $Y_\nu^TY_\nu$ is
real, as can be easily verified.  Thus the light neutrino mass matrix
${\cal M}_\nu$ is real (see Eq. (19)), and there is no CP violation
in neutrino oscillations in this limit, as in the case of Majorana LFV.
We shall use this fit for the case of Dirac LFV case in our numerical
analysis.

\subsection{Predictions for lepton flavor violation}

In our calculation we make the following assumptions about the supersymmetry
breaking sector, adopting the mSUGRA framework.
We assume universal scalar masses $m_0$, a common gaugino mass
$m_{1/2}$ and trilinear soft SUSY breaking $A$ terms that are
proportional to the respective Yukawa coupling matrices, with
a common proportinality constant $A_0$.  We take the fundamental scale
at which these relations hold to be the GUT scale $=2.4 \times 10^{16}$ GeV.
We impose various experimental constraints on the SUSY spectrum.  In
particular, the Higgs boson mass limit ($m_h \geq 114$ GeV) along with the 
$b \rightarrow s+\gamma$ constraint requires that
$m_{1/2} \geq 250$ GeV.  The direct limits on SUSY searches are also
taken into account.  Furthermore, radiative electroweak symmetry breaking
is demanded.  We choose the sign of $\mu$ to be positive, which is
preferred by the $b \rightarrow s+\gamma$ constraint.
We do not explicitly impose constraints from muon $g-2$, but
we note that for the range of parameters we have chosen (positive $\mu$
and moderate to large $\tan\beta$), this anomaly can be explained via
SUSY exchange, provided that $m_{1/2}$ is limited to less than about 800 GeV ($\tan\beta\leq 40$).

\begin{figure}\vspace{-2cm}
\centerline{ \DESepsf(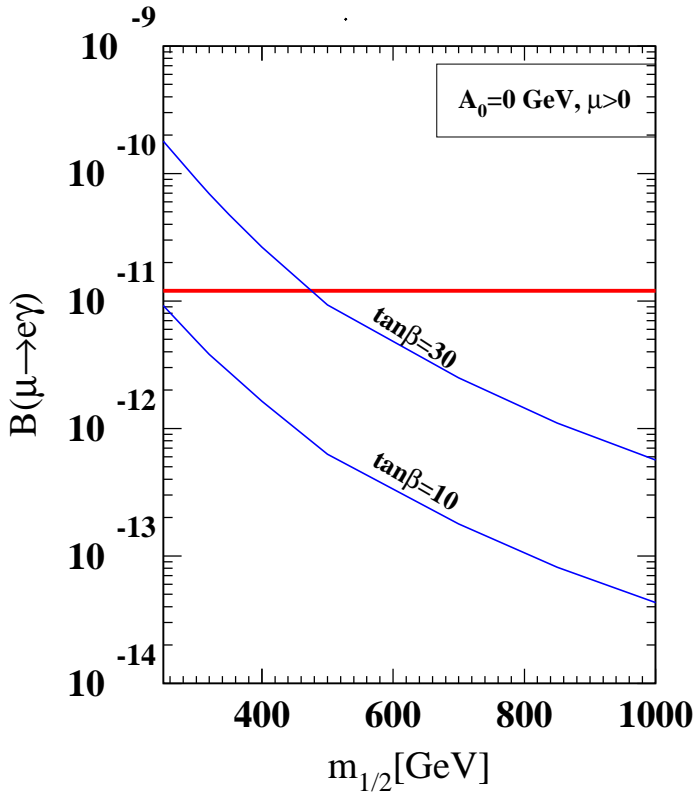 width 8 cm) }
\vspace{0.5cm}\caption {\label{fig2a}$B(\mu\rightarrow
    e+\gamma)$ vs $m_{1/2}$ for two different values of $\tan\beta$.  All
    other parameters are same as in Fig. 1.}
\end{figure}

Additionally,
we require that the  lightest neutralino be the
lightest supersymmetric particle (LSP) in the model.
Furthermore, this LSP should constitute the dark
matter of the universe. However,
for generic SUSY parameters the neutralino annihilation cross section
is not strong enough to bring its relic abundance to the required level
from dark matter constraints
(density in dark matter should be $\sim$30\% \cite{turner} of the critical density in
order for the LSP to form a good dark matter candidate).  This remark
applies for the range of parameters that we are interested in, viz.,
$m_{1/2}\geq 300$ GeV and $\tan\beta \approx 10$.
The dark matter constraint can be satisfied
by invoking the co-annihilation mechanism involving
the LSP and the second lightest SUSY
particle (NLSP), which is the right--handed stau in these models\cite{dark}.
In order for this co-annihilation to be efficient, we  need the
lighter stau mass to be within 25-30 GeV of the LSP mass.
\begin{figure}\vspace{-2cm}
\centerline{ \DESepsf(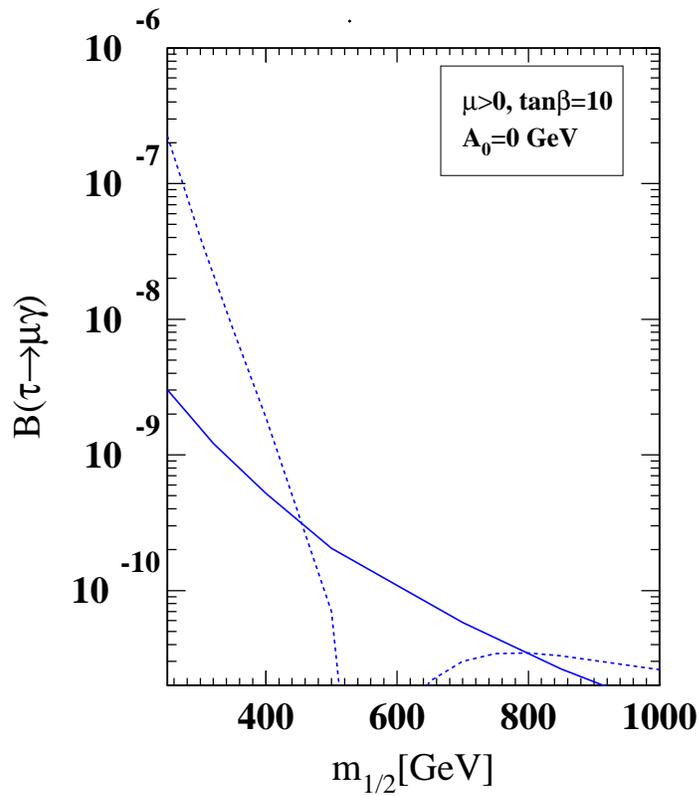 width 8 cm) }
\vspace{0.5cm}\caption {\label{fig2b}$B(\tau\rightarrow
    \mu+\gamma)$ vs $m_{1/2}$ (solid line) for the same set of parameters
    as in Fig. 2 corresponding to the Majorana LFV case.
    The dashed line corresponds to the Dirac LFV alternative given in Eq. (27).}
\end{figure}
We choose the
universal scalar mass $m_0$ such that this constraint
on the lighter stau mass is satisfied.  We allow the range
$0.07<\Omega_{\tilde\chi^0_1} h^2<0.21$. For example, for
the case of $A_0=0$, $m_{1/2} \simeq 4.4 m_0$ will be required
to satisfy this constraint.

We choose $A_\ell = Y_\ell A_0,~A_f = Y_f A_0,~A_\nu = Y_\nu A_0$
at the GUT scale.  For the common $A_0$, we choose three different
values, $A_0 = (0, 300,~800)$ GeV.
We find
that the branching ratio increases with $A_0$, as can be inferred from
the analytic approximation, Eq. (8).

In Fig. 1 we plot the branching ratio for $\mu \rightarrow e + \gamma$
in the Majorana LFV case as a function of $m_{1/2}$, varying it from its
lower limit of 250 GeV to 1 TeV.  Here we chose $A_0=0$.  We find that
the BR varies between $10^{-14}$ and $10^{-9}$.  Part of the parameter
space is already ruled out by the non--observation of $\mu \rightarrow
e+\gamma$.  This experimental lower limit on the BR is indicated by a
thick solid line in Fig. 1.  We conclude that the entire parameter space
can be explored with the planned round of $\mu \rightarrow e+\gamma$
experiment \cite{okada}.

In Fig. 2, we show the $\tan\beta$ dependence of the B$(\mu\rightarrow
e+\gamma)$  for the same set of input parameters as in Fig. 1.
Here we present results for two values of $\tan\beta,
\tan\beta= 10$ and 30. The branching ratio increases with
$\tan\beta$, as anticipated from Eq. (10).
As in Fig. 1, we demand  the dark matter constraint  be satisfied
in Fig. 2 as well.
\begin{figure}\vspace{-2cm}
\centerline{ \DESepsf(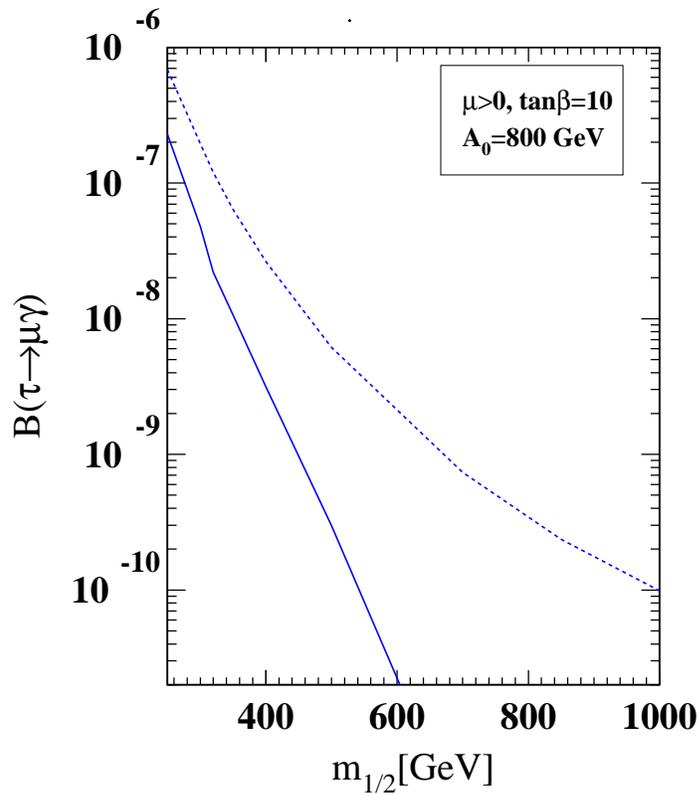 width 8 cm) }
\vspace{0.5cm}\caption {\label{fig3} Same as in Fig. 3, but with $A_0 = 800$ GeV.}
\end{figure}

In Fig. 3 we plot the branching ratio for $\tau \rightarrow \mu +
\gamma$ for the case of $A_0=0$.  The solid line corresponds to the
Majorana LFV case.  For comparison, we also present in the same
graph the $\tau \rightarrow \mu + \gamma$ BR in the case of the
Dirac alternative (fit of Eq. (27)).  In Fig. 4 we plot the same
quantities for a different value of $A_0 = 800$ GeV
and $\tan\beta = 10$.  We find from
Figs. 3 and 4 that the $\tau \rightarrow \mu + \gamma$ is
typically smaller in the Majorana case compared to the Dirac case,
although they can be comparable for some values of $m_{1/2}$ where
there is an accidental cancellation between the chargino and the neutralino
diagrams in the case of Dirac LFV (and not for the Majorana LFV for this
input choice).  The BR for the case of Majorana
LFV is $< 10^{-9}$ for $A_0 = 0$, but can be as large as
$2 \times 10^{-7}$ for $A_0= 800$ GeV.

In Fig. 5, we show the B($\mu\rightarrow e+\gamma$) as a function of
$m_{1/2}$ for the Majorana
alternative (solid lines) and compare it with the Dirac case (dashed line
and dot--sashed line).  We have chosen $A_0=0$ and $\tan\beta=10$
for illustration in this case. The dashed line is drawn for the $Y_{\nu}$ shown
in Eq. (27). We
see that the branching ratio is large in the Dirac
case. However if we scale
down the Dirac neutrino coupling by a factor of 2.5, which does not
alter the light neutrino spectrum (since the overall scale factor  is not
determined from low energy data), we find that the branching ratio
(denoted by the dot--dashed line) decreases.  Thus, depending on the
overall scale factor, $B(\mu \rightarrow e+\gamma)$ can be larger in
the Majorana case compared to the Dirac case (or vice versa).

As Fig. 5 shows, there is some uncertainty in the BR arising from
the overall scale factor.  If we consider ratios of branching ratios,
this uncertainty disappears. We define the ratio
\begin{equation}
r \equiv {B(\mu \rightarrow e+\gamma) \over B(\tau \rightarrow \mu + \gamma)}~
\end{equation}
and study its dependence on the SUSY spectrum as well as on the Dirac/Majorana
cases.  In Figures 6 and 7, we plot $r$
for the cases of  Dirac and  Majorana
alternatives for $A_0=0$ (Fig. 6) and $A_0=800$ GeV (Fig. 7) for $\tan\beta=10$.  We
see from these figures that the ratio depends on the supersymmetry breaking
parameters $A_0$, $m_0$, $m_{1/2}$ etc, in a way that is different for
these two cases.  This difference shows up in almost the entire
parameter space.  We also notice that
the ratio $r$ is large for the Majorana case for large $A_0$.  We conclude that
with
some information on the SUSY breaking parameters, measurement of $r$
can provide crucial insight into the origin of the seesaw mechanism.
\begin{figure}\vspace{-2cm}
    \begin{center}
    \leavevmode
    \epsfysize=8.0cm
    \epsffile[75 160 575 630]{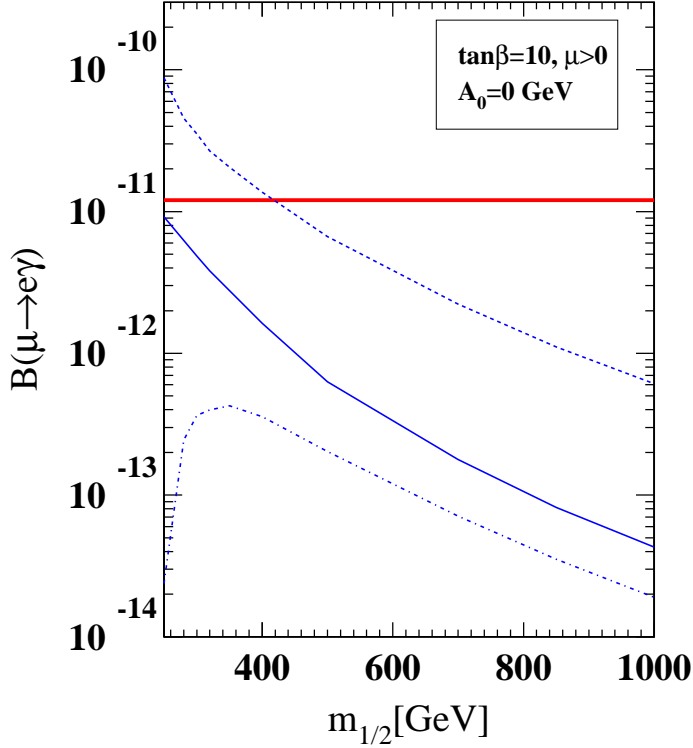}
    \vspace{2.0cm}
    \caption{\label{fig:fig4} $B(\mu\rightarrow
    e+\gamma)$ vs $m_{1/2}$. The solid lines correspond to the Majorana LFV case
    and the dashed line line to the Dirac alternative.  The dot--dahsed line
    is obtained for the Dirac LFV case, but with a rescaling of $Y_\nu$ of
    Eq. (27) by a factor of 2.5.  The horizontal
    solid line shows the current experimental limit.}
\end{center}\end{figure}

\begin{figure}\vspace{-2cm}
    \begin{center}
    \leavevmode
    \epsfysize=8.0cm
    \epsffile[75 160 575 630]{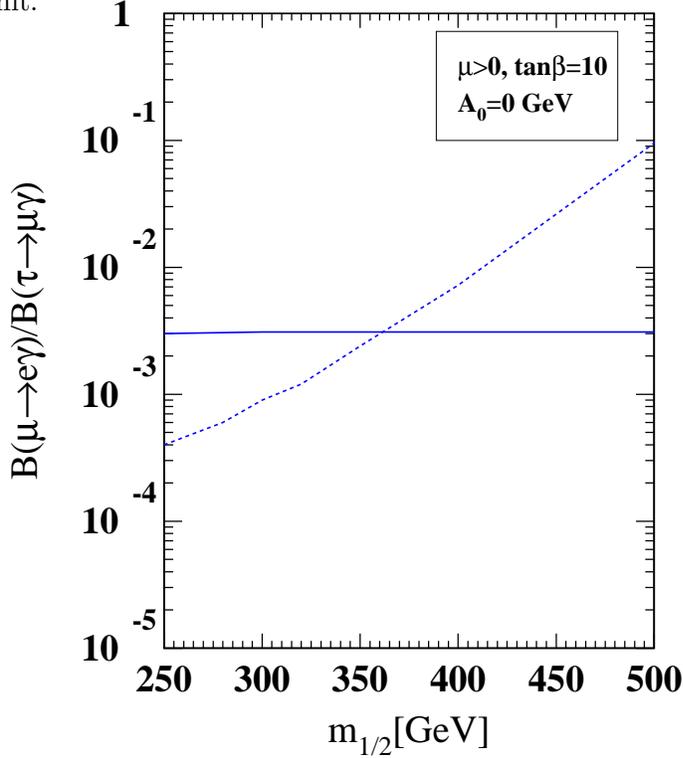}
    \vspace{2.0cm}
    \caption{\label{fig:fig5} The ratio $r=B(\mu\rightarrow
    e+\gamma)/B(\tau\rightarrow\mu+\gamma)$ versus $m_{1/2}$ for $A_0=0$. The solid
    line correspond to the Majorana LFV case
    and the dashed line corresponds to the Dirac LFV alternative.}
    \end{center}
\end{figure}

 \begin{figure}\vspace{-2cm}
    \begin{center}
    \leavevmode
    \epsfysize=8.0cm
    \epsffile[75 160 575 630]{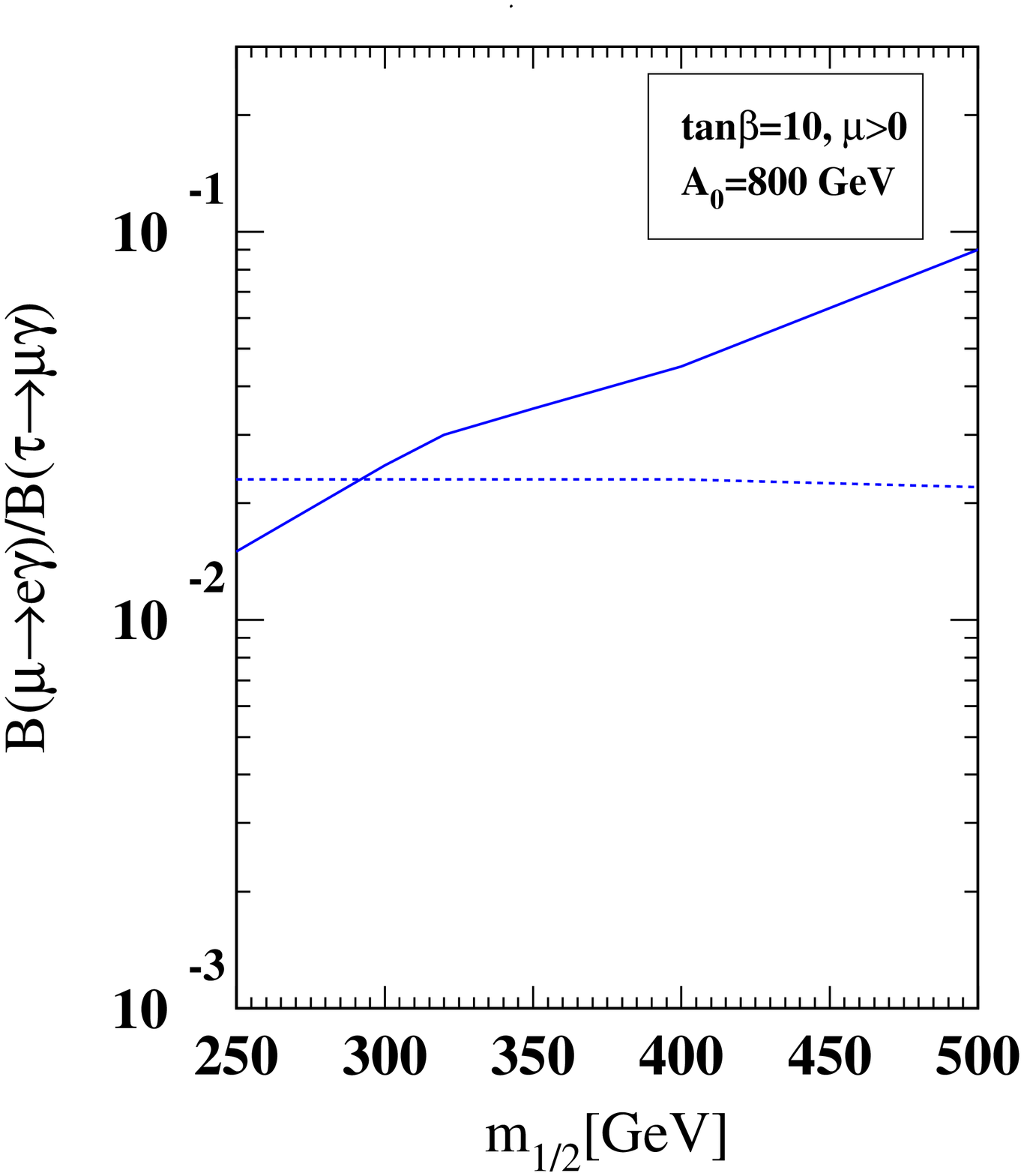}
    \vspace{2.0cm}
    \caption{\label{fig:fig6} Same as in Fig. 6, but with $A_0=800$ GeV.}
\end{center}
\end{figure}

\section{Conclusions}

In this paper we have studied the consequences of implementing the seesaw
mechanism by renormalizable Yukawa couplings involving
a $B-L=2$ Higgs field on lepton flavor violating processes.  The $\nu_R$
Majorana masses arise as  renormalizable Yukawa couplings, rather than
bare mass terms, and therefore influence the renormalization group
evolution of the soft SUSY masses of sleptons.  This
slepton flavor violation is transmitted to the lepton sector and induces
the rare leptonic decays $\mu \rightarrow e+\gamma$ and $\tau \rightarrow
\mu+\gamma$. We have emphasized that this case is qualitatively as well as
quantitatively different from the case where the $\nu_R$ Majorana
masses are directly put in by hand or are induced by non-renormalizable terms.
We have found interesting structure for the Majorana Yukawa coupling matrix
$f$ which is consistent with low energy neutrino oscillation data and that
leads to observable lepton flavor violations in radiative lepton decays.
We have adopted a SUSY
breaking scenario with minimal flavor violation, the popular mSUGRA scheme, and
have found even within this scheme that the radiative lepton decays are within
reach of planned experiments.  The parameter space that we have explored
is consistent with acceptable abundance of neutralino dark matter, as well
as Higgs boson mass limit, $b \rightarrow s + \gamma$ constraint, muon
$g-2$ constraint as well as the direct experimental limits on SUSY
particles.

We have compared the predictions from the Majorana Yukawa coupling related
LFV with that related to the Dirac Yukawa coupling.  The ratio $r=
B(\mu \rightarrow e+\gamma)/B(\tau \rightarrow \mu + \gamma)$, being
insensitive to the overall scale of $B-L$ symmetry breaking,
along with some information on the SUSY spectrum,
can provide
deep insight into the origin of the seesaw mechanism.

\section*{Acknowledgments}

The work of R. N. M. is supported by the National Science Foundation Grant
No. PHY-0099544.  KSB is supported by U.S. Department of Energy grants
DE-FG03-98ER41076 and DE-FG02-01ER45684, and by a grant from the Research
Corporation.

\end{document}